\documentclass{llncs}

\makeatletter
\newcommand{\@chapapp}{\relax}%
\makeatother

\usepackage[title,toc,titletoc,header]{appendix}

\usepackage{tabularx}
\usepackage{graphicx}
\usepackage{subfig}
\usepackage{mathtools}
\usepackage{amssymb}
\usepackage{comment}
\usepackage{algorithmicx}
\usepackage{algpseudocode}
\usepackage{algorithm}
\usepackage{color}
\usepackage{url}
\usepackage[outdir=./]{epstopdf}
\graphicspath{ {./} }
\usepackage{enumerate}
\usepackage{enumitem}
\setlist[enumerate,1]{label=(\arabic*), leftmargin=1cm, labelsep=2mm, topsep=2mm, itemsep=1mm}

\begin{document}

\title{The Complex Link}\author{Hai Zhuge}
\institute{\url{http://www.knowledgegrid.net/~h.zhuge} \\ \email{Email: haizhuge@gmail.com}}
\maketitle

\begin{abstract}
Cyberspace is being more and more tightly linked to the physical space and socioeconomic space to emerge a cyber-physical-socio-economic space, where humans, machines and physical objects in the natural environment interact with each other, efficiently share resources and co-evolve to emerge cyber-physical-social patterns from different spaces. The complex space provides a new reality for experiencing, understanding, thinking and making intelligent decisions. Various links, mappings, and flows bridge and extend the existing understandings to render Cyber-Physical-Socioeconomic Intelligence.
\end{abstract}
\vspace{0.3cm}

\section{Introduction}
\label{sec:introduction}
Computing pioneers envisioned almost all nowadays inventions about cyberspace \cite{engelbart1962augmenting,gray2003next,licklider1960man,turing1950mind}. The challenge of the 21$^{st}$ century computing is moving beyond cyberspace. New requirements like the processing of big data and the development of techniques like various smart machines are pushing forward this movement \cite{hey2009fourth,wilson1989grand}. 
 
From macroscopic, the thought model of the intelligent Future Interconnection Environment based on the Ternary space consisting of the self-evolving physical world, virtual world and mental world proposed in \cite{zhuge2004future,zhuge2005future} has been verified through the development of the Internet of Things (coined in \cite{gershenfeld2004internet} according to Google Scholar) and Cyber-Physical Systems (coined by NSF in 2006).

As the forever challenge, the Future Interconnection Environment has been extended to six spaces: cyberspace, physical space, physiological space, psychological space, socioeconomic space and mental space through various complex links and flows \cite{zhuge2006discovery,zhuge2009communities,zhuge2010interactive,zhuge2010socio,zhuge2012knowledge,zhuge2012probabilistic,zhuge2011semantic,zhuge2016multi}. This movement is changing the way of production, the form of society, and the way of life and thinking. However we still lack theory and methodology to interpret and guide the development. 

New challenge comes from two extremes: the recognition of the nature at the universe scale and at the microscope level, which is verifying and disrupting current cognitions about the nature. 
\vspace{0.3cm}

\section{The Necessity of Linking Human, Machine and the Nature}
\label{sec:necessity}
The development of human society has evolved rich spaces such as mental space, psychological space and economic space, which has significantly changed human behaviors. Studying the complex links between the spaces is an important way to extend human cognition and unveil the reality \cite{zhuge2011semantic}. The winner of Nobel Prize in Economics Richard Thaler successfully built a link between economics and psychology. More and more links and big data collected from the physical space and socioeconomic space through various machines reflects a cyber-physical-socioeconomic space. 

\emph{Can we only study cyberspace by mapping the observed physical space and social space into cyberspace?} The following are the necessity of linking humans, machines and the nature:

\begin{enumerate}
	\item \emph{Direct reflection}. Knowing a small part of the physical space and social space, humans need to constantly explore the laws of the physical space, socioeconomic space and cyberspace with the inspiration from observing and studying these spaces directly. Cyberspace can provide data that facilitate the study of the physical space and socioeconomic space but motivation and original thinking still rely on humans, and the verification of natural science research needs real data from the physical space.
	
	\item \emph{Global view tendency}. Most human related problems are due to lack of global view. Cyberspace reflects only a part of the physical space and socioeconomic space that change constantly, and its capacity relies on the designers of the devices that collect data. Cyberspace is not able to reflect human minds that generate and operate knowledge to develop cyberspace and social space.
	
	\item \emph{Physical trust}. Humans tend to rely on physical objects to recognize and trust things. The following are three instances: people start to recognize and trust a bank with seeing its magnificent building and using its services, people know the taste of a food only after eating it, and people can easily get resident permission if they have houses in the country.
	
	\item \emph{Humanity reference}. Humans tend to rely on communications with other people to trust things through a link like friend. For example, employers usually need references when they recruit people. A reference letter from an authority can significantly help a successful job application.
	
	\item \emph{Authority reference}. Humans do not trust virtual things without the reference to an authorized things endorsed by humans or physical objects. For example, people often do not trust a shop-ping webpage that has not been physically observed and visited before, that has no links from an authorized webpage, or that has not been recommended from friends.
	
	\item \emph{Objective interpretation}. Human minds interpret observations according to various semantic links, especially the cause-effect link. Many experiments show that interpretations are not the reality sometimes, and minds can generate information for interpretation. Experience, belief and characteristics of human sensory organs significantly influence the interpretation. Data sensed by machines can complement the short of human sensory organs. Linking human, machine and the nature can help humans to generate a more objective interpretations.
\end{enumerate}

A one-to-one mapping from simple physical objects or social objects (e.g., individual, relation, and event) into cyber objects can help people to trust cyber objects to a higher extent but mapping complex physical systems and social systems into cyberspace needs real-time simulation. 
\vspace{0.3cm}

\section{Mapping Between Spaces}
\label{sec:mapping}
\begin{figure*}[htb]
	\captionsetup[subfigure]{justification=centering}
	\centering
	\includegraphics[width=\textwidth]{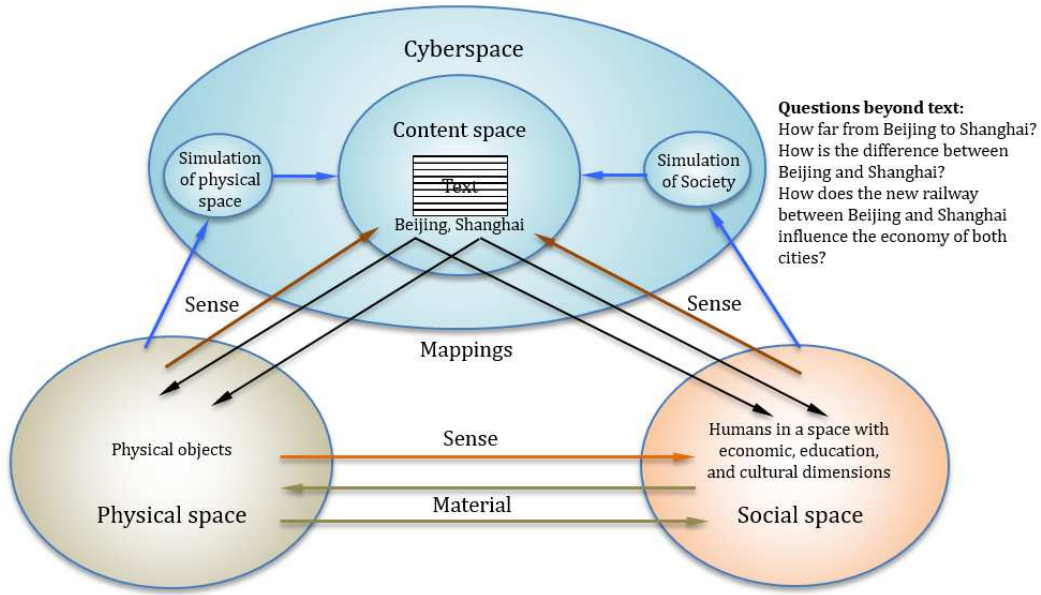}
	\caption{The reality reflected by linking cyberspace, physical space and social space, which enables real-time cyber-physical-social services.}
	\label{fig:linkCPS}
\end{figure*}

Mappings between cyberspace, physical space and socioeconomic space, as depicted in Figure \ref{fig:linkCPS}, enables computing systems to provide cyber-physical-socioeconomic services according to the images in cyberspace, physical space and socioeconomic space. For example, a QA system can answer questions in cyber representations (e.g., texts, images and videos on the Moon), physical representations (e.g., mass) and laws (e.g., Newton’s laws of motion), and socioeconomic representations (e.g., social network and its laws like Metcalfe’s law). These mappings generate the following three images:
\begin{enumerate}
	\item \emph{Cyber Image}, which is built by mapping the physical space and social space into cyberspace through representations of data from sensors (in natural language, formal language, programming languages, or graphical representation).
	
	\item \emph{Physical Image}, which is built by mapping cyberspace into the physical space to identify the physical objects corresponding to cyber images. This mapping carries out through mental behaviors and physical experience with the help of tools like the computing systems with 3D or 4D printers. As a part of material civilization, the physical images evolve with the evolution of humans. All of the physical images constructed by humans form an artifact space. Mapping a text into the physical space enables language representation to get physical images, which is determined by observing and understanding the physical world and representing the understanding in another form of representation just as the use of map in guiding intelligent behaviors. Incorporating this mapping into computing systems enables them to provide services about the physical space. With this mapping, a QA system can answer the questions that do not contain any word within the texts, for example, about the geographical locations of Beijing and Shanghai, the distance between them, and the relation between time, speed and distance. Some physical images have representations in cyberspace, some are in the physical space (e.g., in form of physical models made from materials), and some are only in minds when they have not been represented in cyberspace.
	
	\item \emph{Socioeconomic Image}, which is built by mapping cyberspace into the social space. This mapping carries out through mental behaviors and communication in the social space. Mapping cyberspace into social space enables some words to get social images, e.g., the social image of Beijing is defined in the history of the development of society, including politics, culture, education, economy, population, …, etc. Social image may evolve through time. For example, Beijing determines its general social images as a political, cultural, innovation and international exchange center of China. Some socioeconomic images have representations in cyberspace, some are in form of artifacts, and some are only in minds when they have not been represented in cyberspace or physical space.
\end{enumerate}

Mappings between cyberspace, physical space and socioeconomic space enable a computing system to provide information about the physical space and social space for making a real-time decision with predictive influences on socioeconomic images and physical images. For example, building an express railway between two cities will change the material flow between cities, ecological environment, inhabitant conditions, and economy of two cities. Building a high speed railway (HS2) between London and Birmingham in the UK will stimulates some people living in London to move to Birmingham because of the big differences of living costs and the shortened travel time (travel between two cities will spent the same time as travel within London). This will significantly influence the socioeconomic structure of the two cities, especially the real estate sector.
\vspace{0.3cm}

\section{Optimizing Flows}
\label{sec:optimizing}
Optimizing material flow, information flow, knowledge flow and money flow is a way to realize Cyber-Physical-Socioeconomic Intelligence.
\begin{enumerate}
	\item \emph{The closed loop of material flow} enables materials to be efficiently used by the individual or organization through the loop so that socioeconomic values of the whole network can be increased. For example, industrial ecology is to design and deploy a closed material flow network to enable the waste of one factory to be the production material of another factory in the loop so that the total socioeconomic value of the network can be increased.
	
	\item \emph{The closed loop of information flow} enables every individual in the loop to benefit from an appropriate individual and contribute to another appropriate individual in the loop so that appropriate information can be shared between appropriate individuals. This is a way to avoid information overwhelming. Social medias like Blog and Wechat enable one user to use some operations like “forward” to pass a piece of text to his/her friends who may be interested in the content of the text. This enables information to flow through only the people who need it.
	
	\item \emph{The closed loop of knowledge flow} enables every individual in the loop to learn from an appropriate individual and contribute to another appropriate individual in the loop so that the capacity of every individual can be enhanced. Knowledge emerges through information flow with language (in social media, users can use “comment” operation to express understanding and idea), and knowledge also flows along with questioning and answering processes. Questions and answers may be forwarded as an information flow.
	
	\item \emph{The closed loop of money flow} enables every individual in the loop to earn money from an appropriate individual by providing services or products and spend money by using the services or products provided by another appropriate individual. Money flow usually runs with material flow, information flow and knowledge flow.
\end{enumerate}

Influence through spaces can also form closed loops. Changes in the physical space such as the increase of roads and buildings could influence the existing structure of cyberspace and social space. Changes in the social space like the change of the structure of populations could influence the evolution of the physical space and cyberspace. Changes in the social space and cyberspace could influence the interactions between humans, which influence the development of minds. The mental space evolves with the change of semantic images, which then could influence the behaviors in the cyberspace, physical space and socioeconomic space.
\vspace{0.3cm}

\section{Influence Through Spaces}
\label{sec:influence}
Previous information services such as database query, Web browsing and search, email communication, and social media operate in cyberspace. One operation in cyberspace many influence socioeconomic space or physical space and a change in socioeconomic space or physical space may influence cyberspace through various sensors controlled by programs designed by humans or human direct operations, and operations in cyberspace may influence socioeconomic space and physical space through various actuators.

An email going from the email box of one person to the email box of another person in cyberspace can influence the behaviors of the receiver in the physical space and socioeconomic space, e.g., leading to one person’s traveling for meeting the other person at a hotel (or restaurant) and an event of reaching a cooperation, which may generate socioeconomic influence. The flu propagation in the physical space influences the propagation of the words of the flu used for searching in cyberspace. Operations in cyberspace can be mapped into the physical space and socioeconomic space. The propagation of flu influences public policies and actions of socioeconomic organization. A public policy will generate news and draw attentions in cyberspace and will change human behaviors such as investments in property development, which in turn influence the development of a city, and consequently change the physical space. 

In natural language representation, people often omit the spaces, e.g., an earthquake in the physical space leads to the loss of property and safety in socioeconomic space and people write Blogs in cyberspace to represent what they saw, heard and felt, because humans have established commonsense through common experience and the mappings from language representations into different spaces in minds. People with different knowledge structures can make different mappings. For example, some people who have mathematical knowledge can map a text representation into a mathematical space to form a mathematical model, but those without the knowledge could not make the same mapping. Some representations about the socioeconomic space and the physical space can be easily added but some are difficult to be added. Traditional computing languages omit the space of commonsense (previous study on commonsense are about symbol representation of commonsense). If the cyberspace cannot accurately reflect the evolution of the physical space and socioeconomic space, operations in cyberspace cannot get real influence in the physical space and socioeconomic space.

Researchers in different areas developed different ontologies to help computing systems to make common mappings. However, the current ontologies only interpret common terminologies in some areas but cannot cover human knowledge and thoughts. A natural language query without a space identity can be transformed into an SQL-like query language with a space identity. For example, query the effect of an action \emph{A} in one space \emph{Y} on another space \emph{X} can be represented as following, where * represents the effect of action \emph{A}. 

Query * \emph{in space X} Influenced-By \emph{Action A in space Y}.

The answer can be an action that leads to the change in space \emph{X}. For example, an earthquake in the physical space leads to the action of generating news and the action of forwarding news in cyberspace.

Human actions in different spaces (e.g., physical space and social space) usually cause influence through spaces. Query with the propagation of influence through spaces can take the following form: 

\begin{figure*}[htb]
	\captionsetup[subfigure]{justification=centering}
	\centering
	\includegraphics[width=\textwidth]{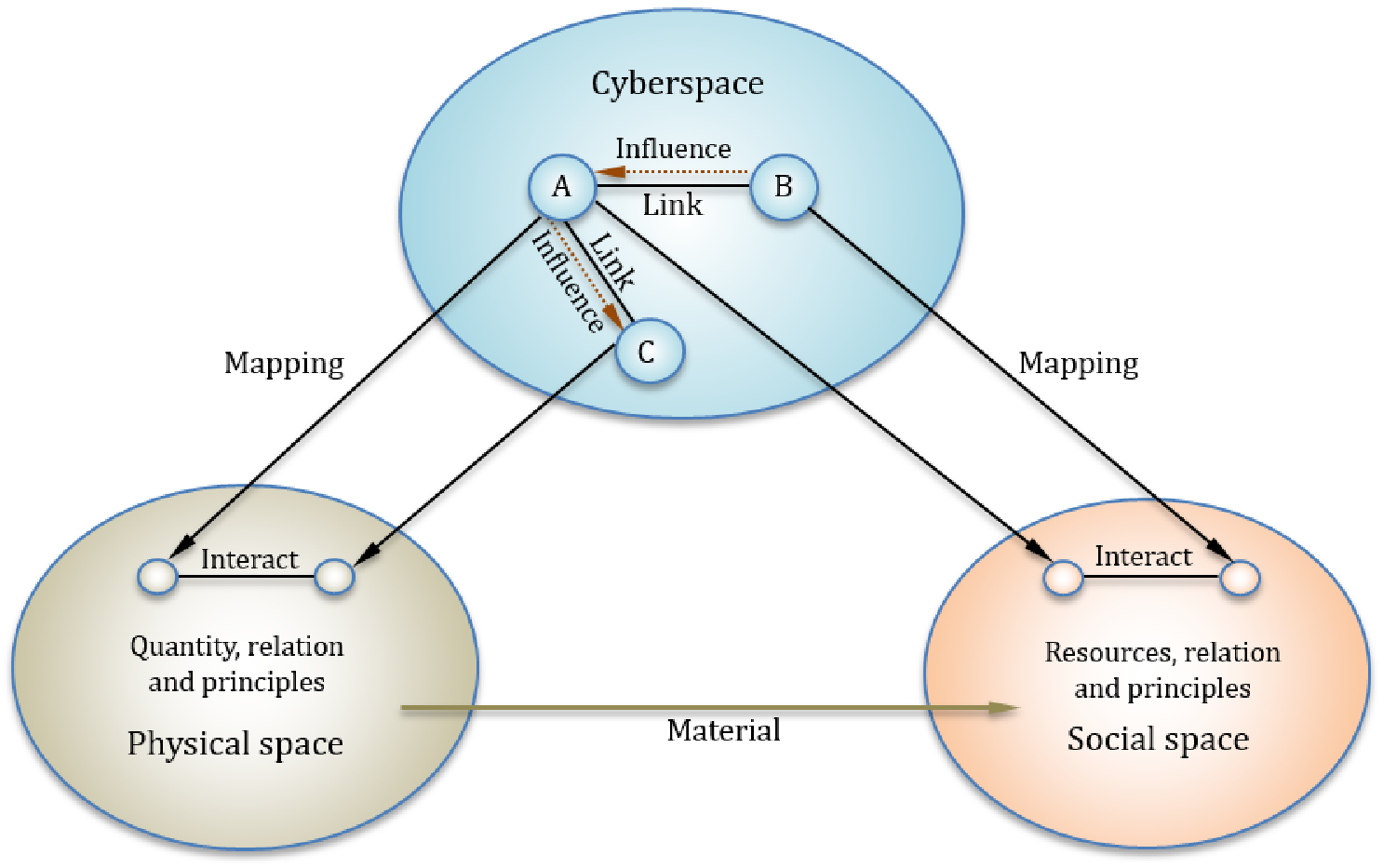}
	\caption{Influence through semantic link alongside interaction in socioeconomic space and interaction in the physical space.}
	\label{fig:influCPS}
\end{figure*}

Query (* \emph{in space Z} Influenced-By \emph{Action B in space Y}) \& (\emph{Action B in space Y} Influenced-By \emph{Action A in space X}). 

Actions in the same space follow some common principles while actions in different spaces may follow different principles, which define different spaces. The quantified prediction of the influence relies on the known principles of the physical space and the principles of the social space, but many principles still need to be discovered in the evolving spaces.

Figure \ref{fig:influCPS} depicts the influence from one individual to the other through semantic link in cyberspace and in the physical space. The influence is generated from the interaction (e.g., competition for resources or reputation) in socioeconomic space following socioeconomic principles or from interacting with each other in the physical space following physical principles. Influence can be either positive or negative. If \emph{A} and \emph{B} share limited resources (e.g., two research students share supervisor’s time) and \emph{B} occupies more resources, then \emph{A} occupies less resources. If \emph{A} has more knowledge and \emph{B} (e.g., \emph{A} is the supervisor of research student \emph{B}), \emph{B} can learn knowledge from \emph{A} (knowledge flows from \emph{A} to \emph{B}).
\vspace{0.3cm}

\section{Cyber-Physical-Socioeconomic Services}
\label{sec:CPSservice}
Linking human, machine and the nature through various complex links supports some distinguished cyber-physical-socioeconomic services about things in different spaces as shown in Table \ref{table:CPSservice}. As all things have life span \cite{zhuge2005future}, the services provided at different times may be different.
\vspace{0.3cm}
\begin{table}
	\centering
	\caption{Cyber-physical-socioeconomic services}
	\label{table:CPSservice}
	\small
	\begin{tabularx}{\textwidth}{|p{1.8cm}|X|p{2.3cm}|p{2.7cm}|p{2.5cm}|}
		\hline
		\textbf{Services} & \textbf{Input / Output} & \textbf{Techniques} & \textbf{Resources}& \textbf{Characteristics}\\
		\hline
		{Search things} &
		{\underline{Input}: A representation of things in cyberspace (including words and images).\newline
		 \underline{Output}: A representation of the things with physical model, and socioeconomic model related to the indicators.} &
		{Identifying things and their spaces.\newline 
		 Modeling things from data collected by sensor network.} &
		{Data on the Web, social medias, and from sensors for things.\newline
		 Social networks of things.} &
		{Provide the things with connections in cyberspace, physical space and socioeconomic space.}\\		
		\hline
		{Discover links} &
		{\underline{Input}: Identity of Things\newline
		 \underline{Output}: Links between things, and the interpretation of the links.} &
	 	{Mining links between things according to frequent patterns, modeling complex things, and relational reasoning.} &
	 	{Web, social medias, and data collected by various sensors in real-time, data about the interested things, and models for complex things.} &
	 	{Provide the links with the influence of adding/removing a link between two things on the values of the other linked things.}\\
	 	\hline
	 	{Recommend things with value analysis} &
	 	{\underline{Input}: User profile\newline
	 	 \underline{Output}: Link between things (people, social organizations, or physical objects). \newline
	 	 					 Predict value change after adding the link.} &
 	 	{Modeling people, object, and organization.\newline Evaluate the value of organization.} &
 	 	{Web, social medias, sensors that can collect real-time data about the interested thing.} &
 	 	{Actively obtain data from physical space or social space according to needs.}\\
		\hline
		{Answer questions about things} &
		{\underline{Input}: Question\newline
		 \underline{Output}: Answer the question with the things about the question and answer.} &
	 	{GIS.\newline Modeling things and society.\newline Social network of things.} &
	 	{Geographical data, Web, social medias, sensors that can collect real-time data about things.} &
	 	{Answer questions beyond cyberspace.}\\
	 	\hline
		{Summarize things} &
		{\underline{Input}: Data about a thing \newline
		 \underline{Output}: A multimedia representation with real-time information on the thing in physical space and socioeconomic space.} &
	 	{Understanding text, picture, video and various languages.} &
	 	{Web, social medias, Wikipedia, and sensors that can collect real-time data about the interested thing.} &
	 	{Summarize things like human expert.}\\
	 	\hline
		{Summarize things} &
		{\underline{Input}: Data about a thing \newline
			\underline{Output}: A multimedia representation with real-time information on the thing in physical space and socioeconomic space.} &
		{Understanding text, picture, video and various languages.} &
		{Web, social medias, Wikipedia, and sensors that can collect real-time data about the interested thing.} &
		{Summarize things like human expert.}\\
		\hline
		{Learn things} &
		{\underline{Input}: \newline Learner’s profile\newline Aim of learning\newline
		 \underline{Output}: The learning process of contents consisting of learning objects in cyberspace, physical space and socioeconomic space.} &
	 	{Linguistics.\newline Educational psychology.\newline Education principle.\newline Physical laws.\newline   
	 	 Motivation.\newline	Value.} &
	 	{Web resources and standards, MOOC, Youtube, social medias, Wikipedia, and sensors that can collect real-time data about things.} &
	 	{Interaction with incorporating relevant physical experience and socioeconomic experience into education process.}\\
	 	\hline
	\end{tabularx}
\end{table}

\section{Cyber-Physical-Socioeconomic Symbiosis}
\label{sec:CPSsymbiosis}
One research issue in industrial ecology is to find and solve environmental problems by analyzing the material flow from the raw materials in the natural environment to the goods through a production process (some materials are transformed into goods to meet the requirements of society while others return to the natural environment as waste). The core of the problem is that the materials flowing back to the natural environment as waste exceed the capacity of the environment. A solution is to redirect the waste material flow to an enterprise as its production materials by connecting the waste flow to an appropriate enterprise as its production materials. The problem is how to find an appropriate enterprise among the existing enterprises or to build a new enterprise that can make use of the materials of waste or minimize the wastes flowing back to the natural environment. However it is difficult for enterprises (especially for small enterprises) to know relevant enterprises within its region and in the world (especially new enterprises) on demand in time.

From the scale of society, materials flowing from the natural environment through various social organizations generate socioeconomic values with a certain waste in macroscopic. From production organization scale, production enterprises are organized by material flows to constitute industrial symbiotic network, where each enterprise only inputs production materials and outputs goods for social consumption or production materials for the other enterprises. The organization can be formed by self-organization of enterprises and recommendation from the social network platform. If enterprises have images in cyberspace, it is possible for cyberspace to find the statuses of the enterprises and control material flows to ensure the minimum waste and the maximum socioeconomic values for sustainable development of society. With the sensors and actuators (including 3D printer), the cyberspace can actively influence the formation of the symbiotic network by finding appropriate links and controlling the material flows according to socioeconomic requirements and the capacity of the natural environment.

The following is the process for designing, building and operating a symbiotic industrial network:

\begin{enumerate}
	\item Determine the method for assessing socioeconomic values of enterprises and their products.
	\item Map enterprises into cyberspace to build their semantic images with input/output flows (including material flow and information flow and their volumes), products (or services), values, and features. A semantic image needs to reflect a physical socioeconomic enterprise from multiple dimensions.
	\item Discover the explicit and implicit links between enterprises by matching the input/output flows, services and features. 
	\item Evaluate the influence of the links and the influence of the flows through links.
	\item Construct a symbiotic network of enterprises according to socioeconomic requirements, the values of the products, the costs of production materials, the cost of production process, the influences of links, and the influence of flows.
	\item Find appropriate enterprises or build new enterprises according to their production materials, products, and estimated socioeconomic values.
	\item Simulate the symbiotic network in cyberspace and adapt the network according to the result of simulation.
	\item Recommend real-time information (including social requirement, market and technology) to enterprises so that they can adapt the material flow to maximize socioeconomic value.
\end{enumerate}	

\begin{figure*}[htb]
	\captionsetup[subfigure]{justification=centering}
	\centering
	\includegraphics[width=\textwidth]{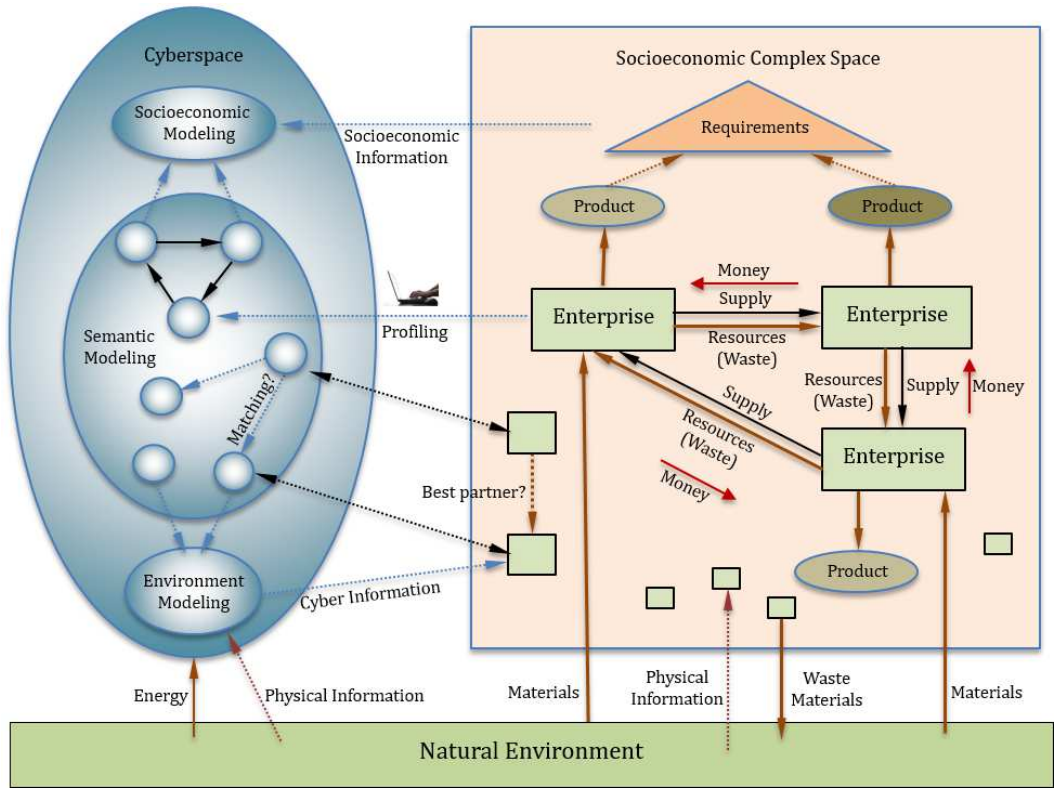}
	\caption{Establishing industrial symbiotic network by semantic modeling, environmental modeling and socioeconomic modeling and closing the loops of material flow and money flow.}
	\label{fig:symbioNet}
\end{figure*}

Figure \ref{fig:symbioNet} depicts a cyber-physical-socioeconomic system, where the cyberspace carries out the semantic modeling of enterprises, physical modeling of the natural environment and social modeling of socioeconomic space, which provide the conditions for determining the appropriate symbiotic link between enterprises and the value for the link and the value for operating the industrial symbiotic network. It also demonstrates the following characteristics of Cyber-Physical-Socioeconomic Intelligence:

\begin{enumerate}
	\item \emph{Cyber-physical-socioeconomic reflexive}. The socioeconomic space reflects the physical space through human minds. Cyberspace reflects enterprises in socioeconomic space and the status of the natural environment, which help the enterprises to make decisions with knowing the situation that is much wider than an individual enterprise. The decisions made by enterprises considering the real-time socioeconomic status. Cyberspace can predict the effect of a decision before making decision and assessing the effect of decision through socioeconomic modeling and environmental modeling. Information flows through physical space, cyberspace and social space, forming and evolving a cyber-physical-socioeconomic space, which reflects more reality than any single space.
	
	\item \emph{Self-adaptive}. The supply link and the flows through the link change with the change of the stages of the lifecycle of enterprises and with the increase of new enterprises based on cyber-physical-socioeconomic reflexive.
	
	\item \emph{Multi-space computing paradigm}, which operates with input flows (material flow, service flow, money flow and information flow) and output flows (material flow, service flow, money flow and data flow). The potentials of enterprises can be discovered to increase their values through global optimization of flows. The key of many problems about human behaviors dues to the unavailability of global information.		
\end{enumerate}

\section{Conclusion}
\label{sec:conclusion}
The existing computing methods are mainly based on modeling the reality. Computing process neglects the reality once a model is built. However models usually reflect a part or a dimension of the reality. Connecting humans, machines and the nature through various complex links enables discovering and formulating problems, finding solutions, and evaluating the solution and its influence in cyberspace, physical space and socioeconomic space, which represents fundamental cyber-physical-socioeconomic intelligence. 
\vspace{0.3cm}

\section*{Acknowledgment}
This work was supported in part by National Science Foundation of China (No. 61640212).
\vspace{0.3cm}

\bibliographystyle{unsrt}
\bibliography{complex_link}

\end{document}